\documentclass[twocolumn,aps,nofootinbib]{revtex4} 
\include{epsf}
\usepackage{graphics}% Include figure files

\newcommand{\be}{\begin{equation}}
\newcommand{\ee}{\end{equation}}
\newcommand{\bea}{\begin{eqnarray}}
\newcommand{\eea}{\end{eqnarray}}

\def\fun#1#2{\lower3.6pt\vbox{\baselineskip0pt\lineskip.9pt
        \ialign{$\mathsurround=0pt#1\hfill##\hfil$\crcr#2\crcr\sim\crcr}}}

\renewcommand\({\left(}
\renewcommand\){\right)}
\renewcommand\[{\left[}
\renewcommand\]{\right]}

\newcommand\eq[1]{Eq.~(\ref{#1})}

\newcommand\GeV{\,\mbox{GeV}}

\newcommand\Mpc{\,\mbox{Mpc}}

\newcommand\msun{M_\odot}
\newcommand\mpl{M_{\rm P}}

\newcommand\lsim{\mathrel{\rlap{\lower4pt\hbox{\hskip1pt$\sim$}}
    \raise1pt\hbox{$<$}}}
\newcommand\gsim{\mathrel{\rlap{\lower4pt\hbox{\hskip1pt$\sim$}}
    \raise1pt\hbox{$>$}}}

\newcommand\diff{\mbox d}

\def\dslash{\not{\hbox{\kern-2pt $\partial$}}}
\def\Dslash{\not{\hbox{\kern-4pt $D$}}}
\def\Oslash{\not{\hbox{\kern-4pt $O$}}}
\def\Qslash{\not{\hbox{\kern-4pt $Q$}}}
\def\pslash{\not{\hbox{\kern-2.3pt $p$}}}
\def\kslash{\not{\hbox{\kern-2.3pt $k$}}}
\def\qslash{\not{\hbox{\kern-2.3pt $q$}}}

 \newtoks\slashfraction
 \slashfraction={.13}
 \def\slash#1{\setbox0\hbox{$ #1 $}
 \setbox0\hbox to \the\slashfraction\wd0{\hss \box0}/\box0 }
 
\def\ee{\end{equation}}
\def\be{\begin{equation}}

\def\calp{{\cal P}}
\def\calr{{\cal R}}
\def\calpr{{\calp_\calr}}

\newcommand\sub[1]{_{\rm #1}}

\newcommand\ncobe{N\sub{COBE}}

\begin{document}
\preprint{DESY 02-136}
\setlength{\unitlength}{1mm}
%\twocolumn[\hsize\textwidth\columnwidth\hsize\csname@twocolumnfalse\endcsname]
\title{New constraints on the running-mass inflation model}
\author{Laura Covi$^1$, David H.~Lyth$^2$ and Alessandro Melchiorri$^3$}
\address{ 
$^1$ DESY Theory Group, Notkestrasse 85, D-22603 Hamburg, Germany.\\
 $^2$ Physics Department, Lancaster University,
Lancaster LA1 4YB, United Kingdom.\\
$^3$ Astrophysics, Denys Wilkinson Building, University of Oxford, 
Keble road, OX1 3RH, Oxford, United Kingdom.}
\date{\today}%
\begin{abstract}
We evaluate new observational constraints on the 
two-parameter scale-dependent spectral index predicted by the
running-mass inflation model by combining the latest Cosmic Microwave
Background (CMB) anisotropy measurements with the recent 2dFGRS 
data on the matter power spectrum, with Lyman $\alpha $ forest
data and finally with theoretical constraints on the reionization
redshift. We  find that present data still allow significant 
scale-dependence of $n$, which occurs in a physically reasonable
regime of parameter space.
\end{abstract}
\bigskip
%\pacs{PACS Numbers: }

\maketitle

\section{Introduction.}

The recent observation of the cosmic microwave background (CMB) 
angular anisotropies~\cite{toco,b97,Netterfield,halverson,lee,cbi,vsa} 
have revealed an outstanding agreement between the data and the 
inflationary predictions of a flat universe and of a primordial 
scale invariant spectrum of adiabatic density perturbations
(see e.g. 
\cite{debe2001,pryke,stompor,wang,cbit,vsat,saralewis,mesilk,caro}). 

Furthermore, CMB measurements are in perfect agreement with
other types of data and a global consistent picture is slowly
emerging. For example, the CMB constraint on the amount of matter 
density in baryons, $\omega_b$, is now consistent with the independent 
constraints from standard big bang nucleosynthesis (BBN)
(see e.g. \cite{hansen}).
Also, the detection of power around the expected third
peak, on arc-minutes scales, provides a new and independent 
evidence for the presence of non-baryonic dark matter
(see e.g. \cite{mesilk}).
At the same time, early data releases from the 2dFGRS and SDSS  
galaxy redshift surveys \cite{2dfgrs,sdss} are living 
up to expectations and combined analysis of all these datasets  
are placing strong constraints on most of the cosmological 
parameters~\cite{efstathiou,lahav,mesilk}.
A combined analysis of both CMB and galaxy clustering is indeed 
necessary in order to break some of the degeneracies between 
the cosmological parameters that are affecting the single datasets.
Recent combined analysis, for example, have provided 
important constraints on the amplitude of the matter 
fluctuations \cite{lahav,mesilk}, on the 
gravity waves background~\cite{efstathiou}, 
on the dark energy equation of state~\cite{bean,percival2,mortsell} 
and on the neutrino mass~\cite{hannestadnu,elgaroy}.

Thank to this new evidence, it is now accepted that the structure 
in the Universe is caused by an almost scale-independent primordial 
density perturbation, existing  before cosmological scales start 
to enter the horizon. Nearly exponential inflation can easily 
explain the origin of this perturbation, because it
converts  the quantum  fluctuation of every light scalar field 
into a scale-independent classical perturbation.
At the present time, there is no equally successful alternative 
explanation for the origin of this perturbation. 
%In particular there is no theory of a bouncing Universe, 
%only of a collapsing ones~\cite{Khoury:2001zk,Brustein:1994kn}.
In particular, there is no reason to expect the quantum-to-classical conversion
to take place in extra space dimensions, though of course a 
string-theoretic description of such dimensions may play a vital 
indirect role by determining the form of the effective four-dimensional 
field theory.

In this paper, we combine the latest CMB and galaxy clustering 
data in order to place constraints on the inflationary model 
responsible for the primordial density perturbation.
We will in fact assume the simple paradigm of a slowly-rolling 
inflaton field, whose perturbations are the seed for the
primordial density perturbations~\cite{treview,book}~\footnote%
{The alternative \cite{curvaton} is to make a  `curvaton' field responsible,  
different from the inflaton if  indeed the latter  exists.}.
If this hypothesis is correct, observation will be able to distinguish 
between the few simple and well-motivated inflation models that exist, 
because they give different predictions for  the spectral index of the 
primordial density perturbation. 
From among such models, we focus in this paper on the running-mass model 
of inflation~\cite{st97,st97bis,clr98,cl98,c98,rs}, 
which is the only one  which can that can give
 a  strongly scale-dependent (running) spectral index. 
Comparing with a suite of observations,
we find that significant running is still allowed.

A similar attempt in constraining this kind of scale-dependence
was made in \cite{cl99} and \cite{covi}, where however only 
a limited subset of the CMB and LSS data then available was used.
A model-independent approach is used instead in \cite{hannestad-run}, 
where constraints in the plane $n$ and its first derivative 
$dn/d\ln k$ were obtained. The main difference with our approach
is that there it is implicitly assumed that the scale dependence 
is small, while this is not necessary the case for the running mass 
model.

%Also we should mention that on the theoretical side efforts have been 
%put in the development of a more general formalism, beyond the 1st 
%order in slow roll, able to describe also the more general scale 
%dependence \cite{leach2}, \cite{kinney02}. 

Our paper is organized as follows: In section II we discuss the 
running mass model. In section III we present our data analysis 
method and results. Finally, in section IV, we discuss our conclusions.

\medskip
\section{The Running Mass Model.}
\medskip

\subsection{The potential}

The running-mass model \cite{st97,st97bis,clr98,cl98,c98,rs} seems 
to be the only one which is both very well-motivated from the
particle physics point of view and also presents a 
strong scale-dependence of the spectral index.
The model is based on supersymmetry and identifies the inflaton 
field with a flat direction of the supersymmetric scalar 
potential. Global supersymmetry is broken during inflation
by the large vacuum energy that drives inflation and so soft 
mass terms are generated for all scalar fields~\cite{rsg}.
Taking the one-loop corrections to the tree-level potential into
account, the most general expression for the potential along
a flat direction, like the inflaton's,
can then be cast into the form
\be
V = V_0 + \frac12m^2(\ln \phi)\phi^2 + \cdots
\,.
\label{runpot}
\ee
where $m^2 $ is the soft inflaton mass and the dots indicate
non-renormalizable terms which are supposed to be negligible because
$\phi/\mpl$ is exponentially small (the usually--dominant
renormalizable quartic term is absent precisely because we are 
using a flat direction in the scalar field space).
The mass dependence on the renormalization scale, in our case
identified with the value of the inflaton field, 
is given by the renormalization group equation  \cite{st97}
\be
{\diff m^2\over \diff\ln\phi} \equiv 
{\diff m^2\over\diff\ln Q} = \beta_m
\label{rge}
\,,
\ee
and therefore is proportional to the inflaton couplings, 
as given below in \eq{betam}. 

Ignoring for the moment the running, a generic supergravity theory 
will generate a soft mass-squared with magnitude at least of order
$|m^2|\sim V_0/\mpl^2$, which is marginally too big to support inflation.
This is a problem for any model of inflation, whether or not the mass 
term is the one that is supposed to dominate. 
In all models except the running mass model, the problem is solved 
either by imposing a global symmetry to protect
the flatness of the potential, or by supposing that the mass-squared 
is accidentally suppressed. The running mass model instead accepts 
the generic value of the mass-squared
at the Planck scale  (or some other high scale), 
and relies on the loop correction to 
sufficiently reduce it in the regime where inflation takes place.

For a particle with gauge and Yukawa interactions, we have
that at one loop $\beta_m $ is given by \cite{st97bis, c98}
\be
\beta_m = - \frac{2 C}\pi \alpha \widetilde m^2
+ {D\over 16\pi^2} |\lambda |^2 m^2_{loop}
\, ,
\label{betam}
\ee
where the first term arises from gauge particles loops and
the second from matter loops.
Above $C, D$ are a positive group-theoretic numbers 
of order one, counting the degrees of freedom present in the loop,  
$\alpha$ is the gauge coupling, and $\widetilde m$ is the gaugino mass,
while $\lambda$ denotes a common Yukawa coupling and  $m^2\sub{loop}$ 
the common susy breaking mass-squared of the scalar particles 
interacting with the inflaton via Yukawa interaction.
Note that the first term in \eq{betam} is always negative,
while the second has no definite sign, since $m^2\sub{loop}$ 
is defined as the mass squared splitting between scalar and 
fermionic superpartners and can have either sign.
Also the case of a very-weakly-interacting inflaton
gives $\beta_m \rightarrow 0$ and so that the constant mass 
potential is recovered.

Over a sufficiently small range of $\phi$, or for
very small inflaton couplings, it is a good approximation to 
take a Taylor expansion of the running mass $m^2(\ln\phi)$ and then
we have:
\be
V=V_0 +\frac12m^2(\ln \bar\phi)\phi^2 
-\frac12 c(\ln \bar \phi) \frac{V_0}{\mpl^2} \phi^2 \ln(\phi/\bar\phi)
\, ,
\label{vlin1}
\ee
where we have expanded around $\bar\phi$, and
rescaled the last term w.r.t. $V_0/\mpl^2$ for future convenience.
The dimensionless constant $c$ is proportional to the 
mass beta function,
\be
c = - \mpl^2 \beta_m/V_0
\label{cofbeta}
\ee

It has been shown \cite{cl98} that for small $c$, as  is required by
slow roll conditions, the linear approximation
is very good over the range of $\phi$ corresponding to horizon exit
for scales between $k\sub{COBE}$ and $8h^{-1}\Mpc$. 
In order to obtain also a crude estimation of the reionization
epoch via the Press-Schechter formula, which involves the power
spectrum at a scale of order $k\sub{reion}^{-1}\sim 10^{-2}\Mpc$ 
(enclosing the relevant mass of order $10^6\msun$), we shall 
assume that the linear approximation is adequate down to this 
`reionization scale'. 

For studying the inflaton potential, it is
very useful to introduce a new parameter $\phi_*$ via
\be
m^2(\bar\phi) = {V_0 \over \mpl^2}\, c(\bar\phi) 
\[\ln\({\phi_*\over \bar\phi}\) +\frac12\]
\, .
\label{mofphi}
\ee
Then \eq{vlin} takes the simple form \cite{cl98}
\be
V=V_0-\frac12 \frac{V_0}{\mpl^2} c\phi^2\( \ln\frac{\phi}{\phi_*}
-\frac12 \) 
\,,
\label{vlin}
\ee
leading to 
\be
\frac{V'}{V_0} = -c\frac{\phi}{\mpl^2} \ln\frac\phi{\phi_*} 
\,,
\ee
In typical cases the linear approximation is valid at $\phi=\phi_*$,
and that point is then a maximum or a minimum of the potential.

The running-mass model supposes that the  relevant soft masses at the 
Planck scale (or some other high scale) have magnitude roughly of order
\be
|\mbox{\rm soft mass}| \sim V_0^\frac12/\mpl \sim H
\,,
\label{massmag}
\ee
and that the relevant couplings (gauge or Yukawa) are very roughly of 
order 1. 
In general, one of the loops will dominate over the others making only
one soft mass relevant besides the inflaton mass. 
With these assumptions, 
the coupling  $c$ defined by \eq{cofbeta} 
is roughly
\be
|c|\sim 10^{-1}\mbox{ to }10^{-2}
\,.
\label{c-range}
\ee
A bigger value of $|c|$ is unviable because it would not allow inflation.
On the other hand, using \eq{vlin}
 as a very crude estimate of the inflaton mass at the
Planck scale, 
 one can see that a much smaller value of $|c|$ is probably not viable
either, since it would 
require the inflaton mass at that scale to be suppressed below
the estimate \eq{massmag}. 

For scalar masses, the estimate
\eq{massmag} is expected to be valid provided that the following statements
are true: (i) there is at most gravitational-strength
 coupling between the inflaton sector and the sector in which
SUSY is spontaneously broken (ii) this breaking comes from
an $F$ term as opposed to a $D$ term (iii) during inflation
(in contrast to the situation in the vacuum)  the
$F$ term in the potential is {\em not} accurately canceled by the other,
negative contribution to the potential. 
The third of these requirements is reasonable provided that the potential
satisfies
\be
V_0\gsim M\sub S^4
\,,
\ee
where $M_S$ is the SUSY-breaking scale in the vacuum.

The most economical assumption is that actually 
the mechanism of spontaneous SUSY breaking during inflation is basically
the same as it is in the vacuum, making
$V_0$ roughly of order $M\sub  S^4$. For respectively the cases
of anomaly-mediated, gravity-mediated and (low-energy)
gauge-mediated transmission of
SUSY-breaking to the Standard Model sector, this leads to 
 the rough estimates
\bea
V_0^\frac14 &\sim & 10^{12}\GeV \\
V_0^\frac14 &\sim & 10^{10}\GeV \\
V_0^\frac14 &\sim & 10^{6} \GeV 
\,.
\eea
The second case was assumed in 
 the original works
\cite{rsg,st97,st97bis}. It gives the correct  normalization for the 
spectrum for reasonable parameters, but there is probably enough latitude
to allow the first or even the third case.

We emphasize that the estimate \eq{massmag} for the
scalar soft masses in the inflaton sector
is reasonable
in all three cases,
even though 
the  soft masses in the Standard Model sector
(evaluated of course in the vacuum, and of order $100\GeV$) 
satisfy that estimate
only in the second, gravity-mediated case.
In the first (anomaly-mediated) case, this is because
no-scale supergravity does not automatically suppress soft scalar
masses during inflation, as it does in the vacuum \cite{treview}.
In the third (gauge-mediated) case it is because one can easily suppose
that the gauge-mediation mechanism simply does not operate in the inflaton
sector.
If the  relevant soft mass is a gaugino mass, the expectation of
\eq{massmag} is not so automatic because gaugino masses can be very 
small with some types of SUSY breaking, but it
is not unreasonable either.

In summary, 
 we expect $c$ in the range~(\ref{c-range})
in a typical running mass model of inflation. As we now see,
this typically leads to significant
 scale-dependence of the spectral index.

\subsection{The spectrum and the  spectral index}

Now we come to the predicted spectrum and spectral index of the
primordial density perturbation. 
We suppose that it is generated by the inflaton field perturbation,
which means that it  is purely adiabatic and gaussian. It is therefore
specified by the curvature perturbation $\calr(k)$,
with $k$ as usual the comoving wavenumber. This quantity is gaussian and
hence specified by its spectrum $\calpr(k)$.

To give the prediction of the running mass model for the spectrum, it is 
convenient to define
\bea
s \equiv  c \ln(\phi_*/\phi\sub{COBE}) 
%= c \ln(\phi_*/\phi\sub{end}) e^{-c N\sub{COBE}} 
\, ,
 \label{sigma}
\eea
where the subscript COBE denotes the epoch of horizon exit for the
COBE scale $k\sub{COBE} \simeq 0.002 (h/\Mpc)$. 

Instead of the  spectrum $\calp_\calr(k)$ we work with the
conventional quantity $\delta_H \equiv \frac25 \calp_\calr^{1/2}$.
 At the COBE scale, the 
prediction of the running-mass model is
\be
\delta_H (k\sub{COBE})= \frac{1}{5 \pi \sqrt{3}} 
\frac{V_0^\frac12}{\mpl \phi_*}
|s|^{-1} \exp(s/c)
\,.
\ee
As already mentioned, the observed value $\delta_H(k\sub{COBE})
=1.9\times 10^{-5}$ is easily obtained, for choices of
$V_0$ and $\phi_*$ that correspond to reasonable particle physics assumptions.

\begin{figure}
\centering
\leavevmode\epsfysize=6.5cm\epsfxsize=6.5cm
\epsfbox{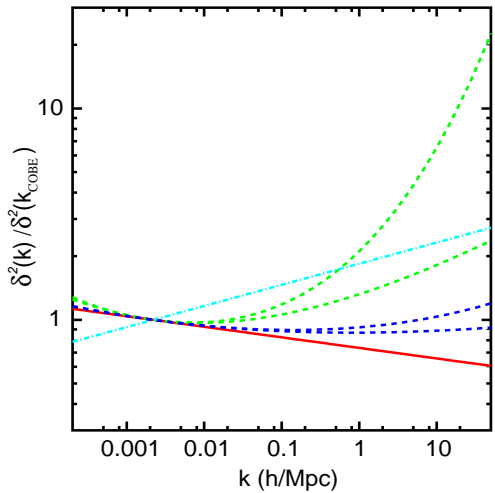}
\caption[delta]{Primordial power spectrum on a logarithmic plot
for the case of a scale invariant spectral index $n = 0.95 $ 
(solid line) and $n=1.1$ (dashed-dotted line), and for the 
running mass prediction with $n\sub{COBE} = 0.95$
and two different $n'\sub{COBE}$, i.e. 
$n'\sub{COBE} = 0.04$, corresponding to 
$c=0.1545, s=0.1295 $ (upper dashed line) or 
$c=-0.1295, s=-0.1545 $ (lower dashed line), 
and $n'\sub{COBE} = 0.01$, corresponding to 
$c=0.08431, s=0.05931 $ (upper dotted line) or 
$c=-0.05931, s=-0.08431 $ (lower dotted line).
}
\label{delta-h}
\end{figure}

In this paper we concern ourselves with the dramatic scale-dependence of the 
spectrum, which is given by 
\be
\frac{\delta^2_H(k)}{\delta^2_H(k\sub{COBE})}
=\exp\[ {2s \over c} \( e^{c \Delta N(k)} -1\)-
2 c \Delta N (k) \]
\,,
\ee
where 
$ \Delta N(k) \equiv N\sub{COBE} - N(k)
\equiv \ln(k/k\sub{COBE}) $.
%counts the number of e-folding from the time when COBE scales
%left the horizon to the end of slow-roll inflation.
We plot the primordial power spectrum for the running mass model
in comparison to the constant spectral index case 
in Figure~\ref{delta-h}.

To discuss the spectral index and its running, we need the
first four derivatives of the potential,
\bea
\mpl^2\frac{V'}{V_0} &=& -c\phi\ln\frac\phi{\phi_*} \\
\mpl^2\frac{V''}{V_0} &=& -c \(  \ln\frac\phi{\phi_*} +1 \) \\
\mpl^2\frac{V'''}{V_0} &=& -\frac c\phi \\
\mpl^2\frac{V''''}{V_0} &=& \frac c{\phi^2}
\eea
Introducing $s$, the first four flatness  parameters \cite{treview,book}
are (setting $V=V_0$ in the denominators)
\bea
\epsilon &\equiv& \frac12 \( \frac{\mpl^2 V''}{V} \)^2
\simeq  { s^2 \phi^2\over \mpl^2} e^{c\Delta N(k)}\\
\eta&\equiv& \frac{\mpl^2 V''}{V} \simeq  s  e^{c\Delta N(k)} -c\\
\xi^2 &\equiv& 
\frac{\mpl^4 V' V'''}{V} \simeq -cs e^{c\Delta N(k)} \\ 
\sigma^3 &\equiv & \frac{\mpl^6 V'^2 V''''}{V} \simeq cs^2  e^{2c\Delta N(k)}
\,.
\label{sigma3}
\eea
The parameters are evaluated at the epoch of horizon exit for the scale
$k$. The first parameter $\epsilon$ is negligible because
 $\phi/\mpl$ is taken to be exponentially small. The condition for slow-roll
inflation is therefore just $|\eta|\ll 1$, which is satisfied in the
regime  $\phi \sim \phi_*$ provided that $|c|\ll 1$. This corresponds
also to requiring $|s|\ll 1$.

Additional and generally stronger constraints on $s$ follow from the 
reasonable assumptions that the mass continues to run to the end of 
slow-roll inflation, and that the linear approximation remains  
{\em roughly} valid. 
Discounting the possibility that 
 the end of inflation is very fine-tuned, to occur close to the
maximum or minimum of the potential, we have the lower bound 
\be
|s|\gsim e^{-c\ncobe}|c|
\,.
\label{sb1}
\ee
Note that for negative $c$, this constraint is very strong,
requiring a very large value of $s$ even for small $c$
and a kind of fine-tuning between $s$ and $c$ to give a 
reasonable value of $n-1$. 

\begin{figure}
\centering
\leavevmode\epsfysize=6.5cm \epsfbox{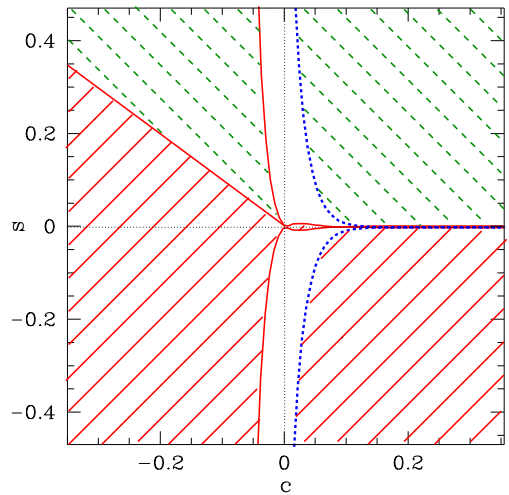}
\caption[]{Theoretical expected valued of the parameters $c, s$
for $N\sub{COBE}=50$; the solid-line-hatched region is strongly 
excluded by naturality assumptions, while the dashed-line-hatched 
region is only weakly excluded. The dotted line shows the prediction 
for the simplest case, $s = e^{-c N\sub{COBE}}$, 
where the linear approximation is valid up to the end of inflation,
triggered by $\eta =1$.
}
\label{laura101}
\end{figure}

For positive $c$, we also obtain a significant upper bound
by setting $\Delta N=\ncobe$ in \eq{runpred}, and remembering
that slow-roll requires $|n-1|\lsim 1$:
\be
|s|\lsim e^{-cN\sub{COBE}}\hspace{2em}(c>0)
\,.
\label{sb2}
\ee
In the simplest case, that slow-roll inflation 
ends when $n-1$ actually becomes of order 1, this bound becomes
an actual estimate, $|s|\sim e^{-cN\sub{COBE}}$.
As discussed in \cite{cl99}, this upper bound can be relaxed for
positive $s$ if the running of the mass ceases before the end
of slow-roll inflation.
The approximate region of the $s$ versus $c$ plane excluded by these
considerations is shown in Figure~\ref{laura101}.

Since $\epsilon$ is negligible, the spectral index is
\bea
n(k) &=& 1 + 2\eta -6\epsilon \\
&\simeq & 1 + 2\eta 
\,.
\eea
This gives
 \cite{cl99}
\be
{n(k)-1\over 2} = 
s e^{c\Delta N(k)} - c \label{runpred}
\,.
\ee
The first derivative of the spectral index is given by
\be
n'(k) \equiv \frac{\diff n (k)}{\diff \ln k} 
= 2 s c\, e^{c \Delta N(k)}
\,.
\label{diffn}
\ee
Clearly the spectral index is not constant within cosmological 
unless $s$ or $c$ is very close to zero.
%scales, if $s$ or $c \neq 0$. 

At the 
%lowest order, at 
COBE scale,
\bea
n\sub{COBE}-1 &=& 2 (s-c) \\
n'\sub{COBE} &=&  2sc
\label{n-dn}
\,.
\label{ncobe}
\eea
The straight line $s=c$ in the $s$ vs $c$ plane
corresponds to $n\sub{COBE}= 1$. The scale independent case 
of constant $n=1$ is given by the origin $s=c=0$, while 
constant spectral index different from $1$ is realized either
on the $c=0$ axis for $s = (n-1)/2$ or on the $s=0$ axis
for $c = - (n-1)/2$.

Taking cosmological scales to span a range $\Delta N\sim 10$, we see
that the running mass model generates a change $\delta n \sim 10sc$
in the spectral index. This change should eventually be
detectable unless both $|c|$ and $|s|$ are towards the bottom of their
expected ranges.

We will show in the following the allowed region both in the
$s, c$ parameter space and in the $n'\sub{COBE} $ 
vs $n\sub{COBE}-1$ plane. 
Note in this respect that not all values of $n'\sub{COBE} $ 
are allowed in the running mass models: from the definition above,
we obtain the constraint
\be
n'\sub{COBE} \geq - {(n\sub{COBE}-1)^2\over 4}
\, ,
\label{bound-dndlnk}
\ee
so that a decreasing spectral index is possible only if $n\sub{COBE}$
is different from 1. 
Also \eq{n-dn} is symmetric under reflection with respect to 
the $s+c =0$ line, so fixing
the variables $n\sub{COBE}-1$ and $n'\sub{COBE} $ 
determines two values of $s, c$. We will investigate in the following
how strongly the data are sensitive to a variation of $n'$
and therefore able to disentangle this degeneracy.
From all this, it follows that fitting for arbitrary value of $n'$
is not exactly equivalent to performing a fit for the running mass 
models.

These predictions  for the spectral index and its derivative
are to leading order in slow-roll. To second  order, the spectral
index is given to a good approximation by \cite{treview}
\be
{n(k)-1\over 2} = 
s(1+1.06c)  e^{c\Delta N(k)} - c 
\label{runpred2}
\,,
\ee
We  neglected a term of order $\eta^2$, and the remaining (unknown)
 error is expected to be of order $\sigma^3$. 
Both of these are negligible which means that
the second-order correction  should be rather accurate.
We shall not include it though, because it just corresponds to a 
small rescaling of the quantity $s$, whose precise value depends on 
unknown physics \cite{treview}. Note that it is in any case easy
to obtain the results at next-to-leading order, by rescaling $s$
by the factor $1/( 1+1.06 c)$ or shift $n\sub{COBE}-1$
by $1.06\, n'\sub{COBE} $.

\medskip
\section{Observational constraints}
\medskip

\subsection{Method}

We compare the recent cosmological observations with a grid
of theoretical models computed with the public available
CMBFAST \cite{sz} code.
We restrict our analysis to flat, adiabatic, 
$\Lambda$-CDM models and we
sample the various parameters as follows:
$\Omega_{cdm}h^2\equiv \omega_{cdm}= 0.01,...0.40$, in steps of  $0.01$;  
$\Omega_{b}h^2\equiv\omega_{b} = 0.001, ...,0.040$, 
in steps of  $0.001$ and $\Omega_{\Lambda}=0.0, ..., 0.95$, 
in steps of  $0.05$. 
The value of the Hubble constant is given by
$h=\sqrt{{\omega_{cdm}+\omega_b} \over {1-\Omega_{\Lambda}}}$

We allow for a reionization of the intergalactic medium by
varying also the compton optical depth parameter 
$\tau_c$ in the range $\tau_c=0.0,...,0.20$ in steps of $0.05$.
Greater values of $\tau_c$ are in disagreement with recent estimates of 
the redshift of reionization $z_R\sim 6 \pm 1$ (see e.g. \cite{gnedin})
which points towards $\tau_c \sim 0.05$.
As discussed later, given a cosmological model, 
we checked for the consistency of the assumed optical depth with its 
expected values derived from a simple reionization scenario. 

We vary the $2$ parameters of the running mass model in the range
$-0.30 < c <0.30$ and $-0.47 < s < 0.47$ in a $50 \times 50$ grid.
To save computing time, we compute 
the CMB anisotropies transfer functions for each cosmological
model just one time and then we integrate them different times
looping over $c$ and $s$.

For the CMB data, we use the recent results from the 
BOOMERanG-98, DASI, MAXIMA-1,CBI, and VSA experiments. 
The power spectra from these experiments were estimated in  
$19$, $9$, $13$, $14$ and $10$ bins respectively
(for the CBI, we use the data from the MOSAIC configuration), 
spanning the range $2 \le \ell \le 3500$.
However, since in this work we are interested only in
comparing the data with the expected {\it primary} anisotropies, 
we limit our analysis to the region $\ell < 1500$ which is likely
not to be affected by secondary effects like 
Sunyaev-Zel'dovich (see e.g. \cite{nabila}). 

The likelihood for a given theoretical model is defined by 
 $-2{\rm ln} {\cal L}=(C_B^{th}-C_B^{ex})M_{BB'}(C_{B'}^{th}-C_{B'}^{ex})$ 
where  $M_{BB'}$ is the Gaussian curvature of the likelihood  
matrix at the peak and $C_B$ are the experimental (theoretical)
band powers.
We discard the first bin of the CBI dataset ($0 < \ell < 400$), 
due to the asymmetric window function and the high sample variance.

We consider $10 \%$, $4 \%$, $5 \%$, $3.5 \%$  and $5 \%$ 
Gaussian distributed  
calibration errors for the BOOMERanG-98, DASI, MAXIMA-1, VSA, and CBI 
experiments respectively and we include the beam uncertainties 
by the analytical marginalization 
method presented in \cite{sara}.

In addition to  the CMB data we incorporate the real-space power spectrum  
of galaxies in the 2dF 100k galaxy redshift survey using the 
data and window functions of the analysis of Tegmark et al. \cite{thx}. 
 
To compute ${\cal L}^{2dF}$, we evaluate $p_i = P(k_i)$,  
where $P(k)$ is the theoretical matter power spectrum  
and $k_i$ are the $49$ k-values of the measurements in \cite{thx}.  
Therefore we have $-2ln{\cal L}^{2dF} = \sum_i [P_i-(Wp)_i]^2/dP_i^2$, 
where $P_i$ and $dP_i$ are the measurements and corresponding error bars 
and $W$ is the reported $27 \times 49$ window matrix. 
We restrict the analysis to a range of scales where the fluctuations 
are assumed to be in the linear regime ($k < 0.2 h^{-1}\rm Mpc$). 
When combining with the CMB data, we marginalize over a bias $b$  
considered to be an additional free parameter. 

We also include the recent $13$ data points on the matter power spectrum 
obtained by Croft et al. \cite{croft} using Lyman $\alpha$ forest
data from $53$ quasar spectra.
The theoretical transfer functions are in this case
computed up to redshift $z=2.72$ and compared by a 
simple chi-square analysis with the data, including the $\pm 25 \%$ error
in the data amplitude.

Moreover, following a similar approach used in a previous analysis
\cite{covi} we use reionization bounds.
Practically, we deduce for each theoretical model, the expected value 
of $\tau_c^{th}$,
\be
\tau_c^{th} = {0.035\,\Omega_b h\over 1-\Omega_\Lambda}
\(\sqrt{(1-\Omega_\Lambda) (1+z_R)^3 + \Omega_\Lambda} -1 \)
\ee
from the reionization redshift $z_R$ estimated 
using a Press-Schechter formula. Taking $f$ to be the
fraction of matter collapsed into objects with mass
$M = 10^6 \msun$, $1+z_R$ is given by 
\be
1+z_R \simeq \left\{
\begin{array}{lr}
{\sqrt{2} \sigma(M) \over \delta_c g(1-\Omega_\Lambda) } 
\mbox{erfc}^{-1}(f) & 
\hspace*{1cm} f \ll 1 \\
{\sigma(M) \over g(1-\Omega_\Lambda) } & 
f \simeq 1
\end{array}
\right.
\ee
where $\sigma (M) $ is the present, linearly evolved, rms density contrast
with top-hat smoothing and $\delta_c =1.7$ is the overdensity contrast
required for gravitational collapse; $g(1-\Omega_\Lambda) $ is the 
suppression factor of $\sigma (M)$ at the present epoch for the case
of a non vanishing cosmological constant:
\bea
g(\Omega) = {5\over 2} \Omega \({1\over 70} +
{209\over 140} \Omega - {1\over 140} \Omega^2 + \Omega^{4/7} \)^{-1}\, .
\eea

\begin{figure}
\centering
\leavevmode\epsfysize=6.5cm \epsfbox{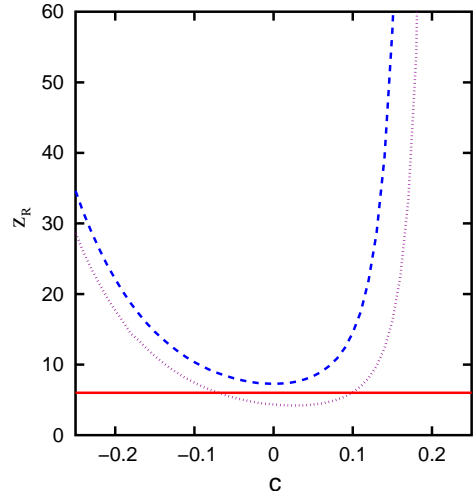}
\caption[]{Theoretical expected valued of $z_R$ for $f\simeq 1$
as a function of $c$ for the case $s=c$ (dashed curve) and 
$s=c-0.05$ (dot-dashed curve).
The cosmological parameters are chosen as $h=0.72$, 
$\omega_{cdm}=0.086 $ and $\omega_b = 0.02$.
We show also the reference line $z_R = 6$ (solid line).
}
\label{zR-c}
\end{figure}

In the case of the running mass model, $\sigma (M)$ can become
very large even for moderate values of $c$, \linebreak
if $c \ln(k_R/k_{COBE}) \simeq 1 $, so that the reionization constraint 
is very powerful.
To give a feeling of this strong dependence, we show in Figure~\ref{zR-c} 
the value of $z_R$ as a function of $c$ for the simple cases $s=c,
s=c-0.05$ at fixed cosmological parameters.

For each model in the database with a given optical depth 
$\tau_c$ we then consider the further prior 
$2ln{\cal L}^{rei} = [\tau_c-\tau_c^{th}]^2/0.05^2$ where the
$1-\sigma$ error should take into account the uncertainties
due to the different reionization mechanisms.

Finally, we also include a constraint on the variance $\sigma$ on 
a mass scale of $10^6 h^{-1} \msun$ of $9\pm1$ consistent
with the limits on the linear power spectrum obtained using
measurements of substructure in gravitational lens
galaxies in the analysis of \cite{dalal}.

One should keep in mind that there are many caveats in the inclusion
of all above constraints. 

We attribute a likelihood  
to each value of $c$ and $s$ by marginalizing over the 
{\it nuisance} parameters.  
We then define our $68\%$ ($95 \%$),  
confidence levels to be where the integral of the  
likelihood is $0.16$ ($0.025$) and $0.84$  
($0.975$) of the total value. 
 
\medskip
\subsection{Results}
\medskip

\begin{figure}
%\centering
\parbox{7.5cm}{
\leavevmode\epsfysize=6.5cm \epsfbox{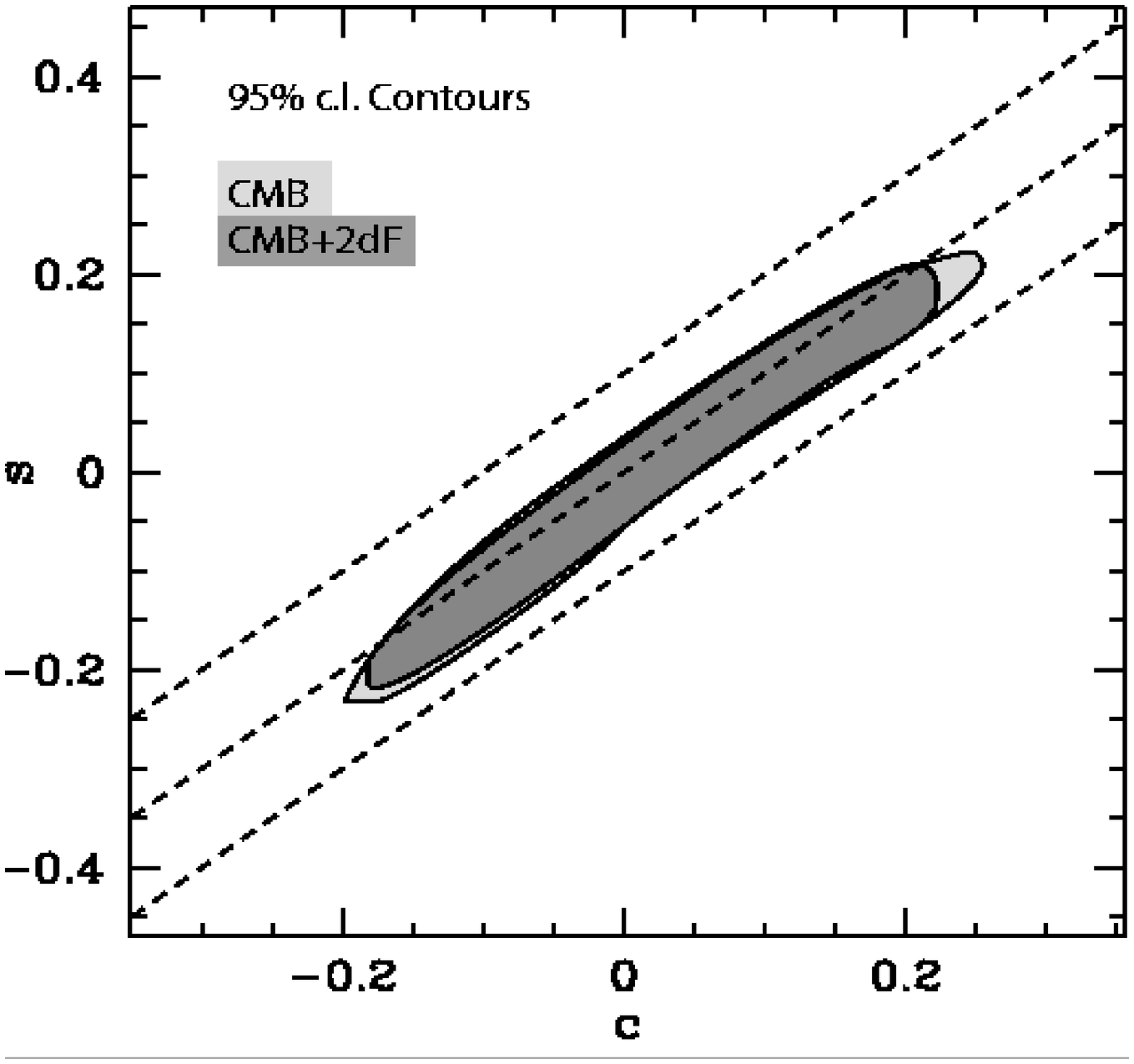}\\
\leavevmode\epsfysize=6.5cm \epsfbox{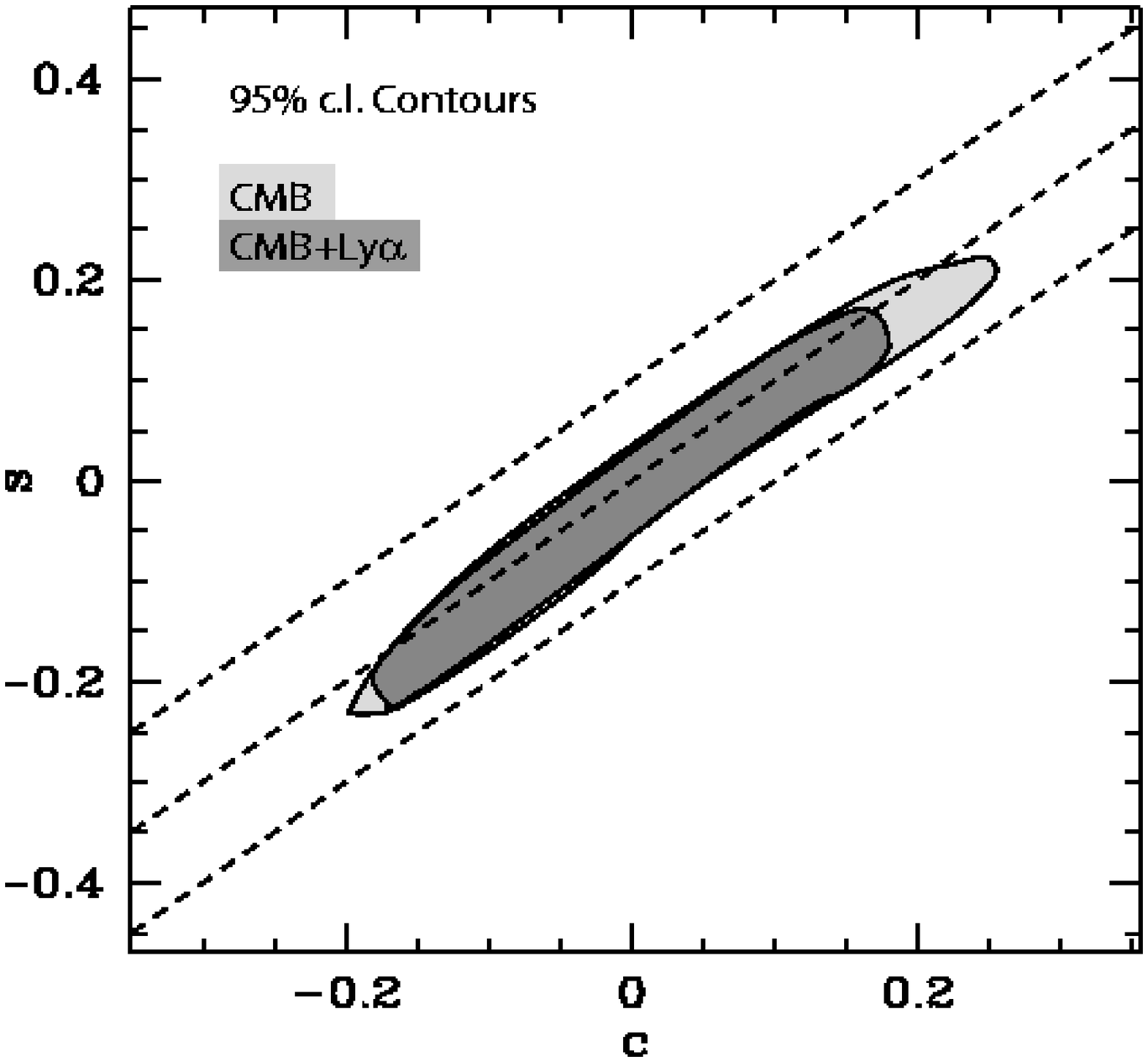}\\
\leavevmode\epsfysize=6.5cm \epsfbox{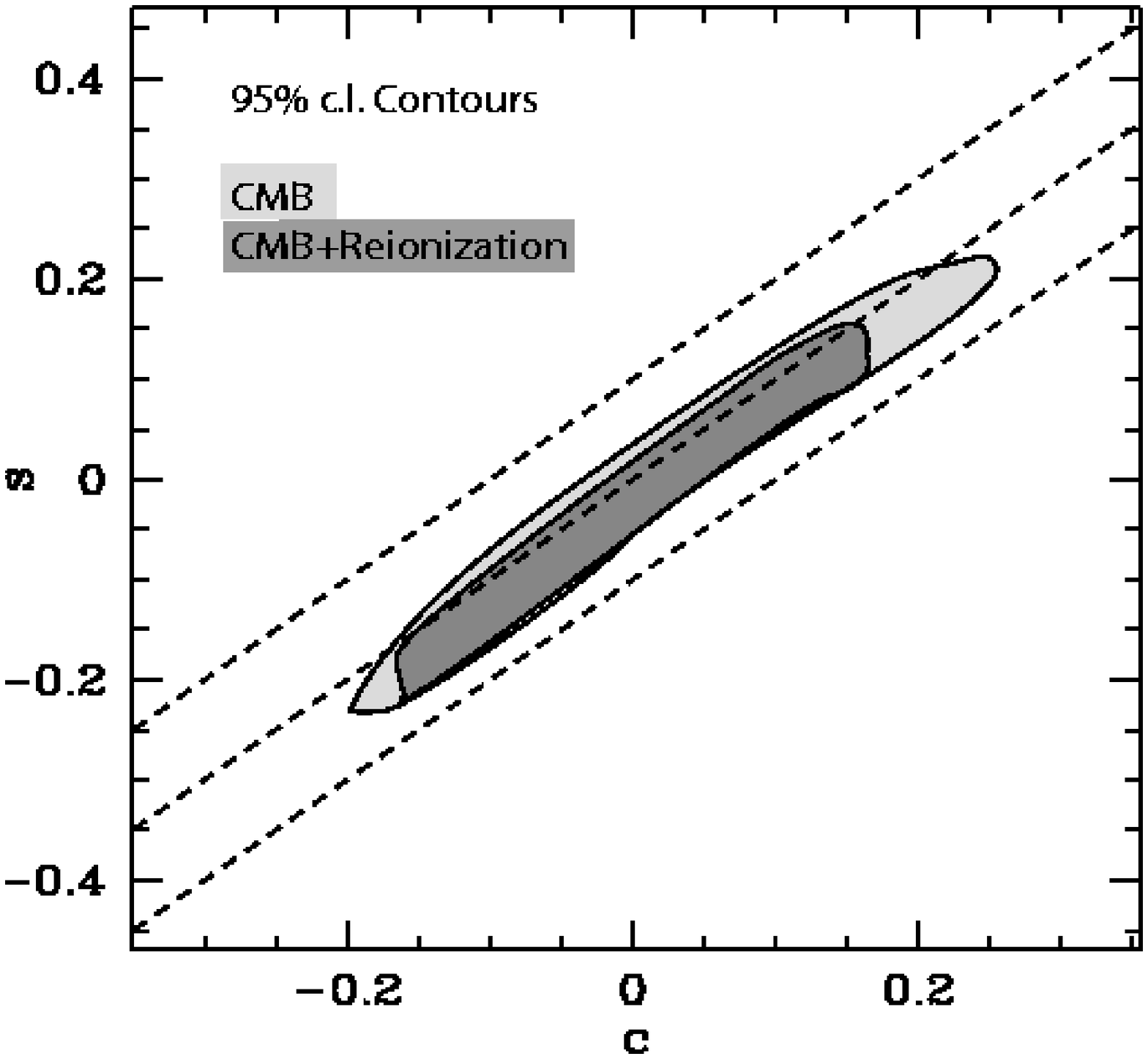}
}
\caption[sc-fig1]{Likelihood contours in the $c-s$ plane.
The gray contours are the $95 \%$ c.l. from the CMB analysis. 
The dark contours are from CMB+2dF analysis (Top), CMB+Lya (Center Panel),
CMB+Reionization constraint (Bottom Panel).The straight lines corresponds to 
$n_{COBE}=0.8$,$1.0$ and $1.2$. }
\label{contours}
\end{figure}

The likelihood contours in the $c-s$ plane 
obtained analyzing the CMB data under the assumption of a set of
``weak'' priors ($h=0.65\pm0.2$, $t_0 > 11 Gyrs$)
are plotted in Figure ~\ref{contours}.
As we can see, even if the space of models analyzed is quite broad,
the CMB data is able to give strong constraints along the 
$s-c$ direction. In particular, we obtain the constraint
$n_{COBE}=1+2(s-c)=0.96_{-0.04}^{+0.06}$ and $n_8=  $ 
at $68 \%$ c.l...
We found that including stronger priors 
($h=0.72\pm0.08$, $\Omega_m=0.3 \pm 0.1$) on the nuisance parameters
does not change these results and does not improve significantly 
the constraints on $n_{COBE}$.
Assuming negligible reionization ($\tau=0$) gives 
$n_{COBE}=0.93_{-0.03}^{+0.03}$.

Also plotted in Figure~\ref{contours} (Top Panel) 
are the likelihood contours obtained in the combined 2dFGRS-CMB analysis.
As we can see, even the inclusion of the 2dF data does not improve 
significantly the CMB constraints.
The combination of the 2dF data is extremely powerful in 
constraining parameters like $\Omega_m$ and $h$, that define the
epoch of equality and the position in the $k$ space of the
matter spectrum. However, in the CMB+2dF combined analysis, we found 
$n_{COBE}=0.97_{-0.04}^{+0.05}$ at $68 \%$ c.l..

In the center panel of Figure~\ref{contours} we report
 the likelihood contours obtained in the combined CMB+Ly-$\alpha$ analysis.
As already noticed by \cite{hannestad-run}, we find that the 
Lyman $\alpha$ data are able to restrict more strongly the scale 
dependence of the spectral index and therefore exclude the
parameter space at large $|c|$; in particular at $68 \%$ we have
$ sc < 0.026 $, so that a variation of the spectral index 
over cosmological scale of the order of 10\% is allowed.

Finally, in the bottom panel of Figure~\ref{contours} we plot the 
likelihood contours
obtained in the combined CMB+reionization analysis.
Also in this case part of the region of large $s,c$ is excluded,
but still a sizable variation and values of $|c|$ up to 0.2
are allowed.
%[{\tt extra words}] 
It should be emphasized that this reionization
constraint is only one of theoretical consistency. As one can see from
Figure~\ref{zR-c},
%[{\tt if you decide to include that plot of $z_R$ versus $c$ Laura}])
the self-consistent value of $z_R$ becomes very large for $c> 0.10$. 
If it turns out that $z_R\lsim 10$,  the top part of the allowed 
region `CMB $+$ Reionization' will be excluded.

\begin{figure}
\centering
\epsfysize=5.5cm 
\epsffile{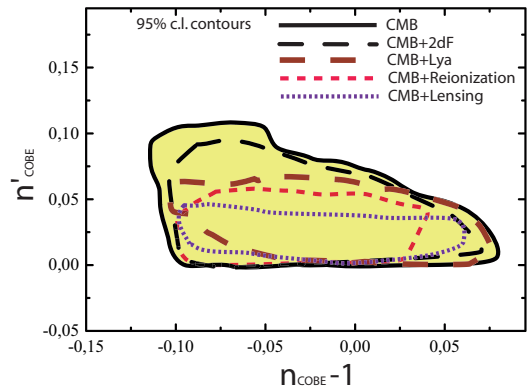}
\caption[sc-fig1]{Likelihood contours in the 
$(n_{COBE}-1)-(n'\sub{COBE})$ plane
for different datasets.
}
\label{contours2}
\end{figure}

Focussing on the region ($s+c>0$ , $c>0$),
%[{\tt alessandro, I think this is what you are plotting}]
we plot the likelihood 
contours in the $2(s-c)$ vs. $2sc$ plane in Figure~\ref{contours2}. 
As we explained before, $2(s-c)$ gives the value of the spectral 
index $n_{COBE}-1$ on COBE scales, while $2sc= n'\sub{COBE} $ gives 
the bend in the spectrum.

As we can see from the figure, the recent CMB data alone constrains
$n'\sub{COBE} <0.1$ at $95 \%$ C.L., while including the constraints from 
2dF, Ly-$\alpha$, reionization redshift, limits the amount 
of deviation from scale invariance to 
$n'\sub{COBE} <0.05$ at $95 \%$ c.l..
Adding the strong lensing constraint from \cite{dalal},
 as discussed earlier, further constrains $n'<0.04$ at $95\%$ c.l..
It is important to notice that a lower spectral index 
$n <0.95$ is consistent with the lensing observation only
if $n'\sub{COBE}>0$. Note also that in all cases the best fit
value of $n'\sub{COBE}$ is not in the centre of the allowed
region, but lower, towards the value $n'\sub{COBE}=0$. 
Furthermore, the inclusion of a bend in the power spectrum,
does not seem to affect in a considerable way the CMB 
constraint on $n$ since there is a weak correlation between
these two variables.

\begin{figure}
\centering
\epsfysize=15cm 
\epsffile{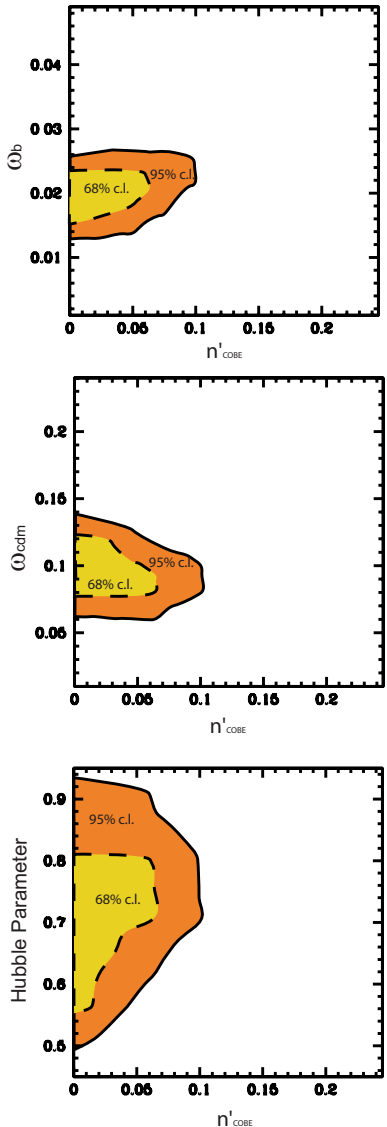} 
\caption[sc-fig2]{Correlations between $n'\sub{COBE}$ 
with different
parameters. From top to bottom, we consider the physical
density in baryons $\omega_b$, in cold dark matter $\omega_{cdm}$
and the Hubble parameter, $h$.
}
\label{contours3}
\end{figure}

The correlation between $n'\sub{COBE}$ and the remaining
parameters is investigated in Figure~\ref{contours3} where
the likelihood function for $n'\sub{COBE}$ 
is plotted in function of $\omega_b$, $\omega_{cdm}$ and 
the Hubble parameter, $h$.

As we can see, the correlation with these parameters is extremely small,
meaning that the inclusion of a bending in the primordial
spectrum would not drastically affect the actual 
constraints on the parameters considered.
However, an increase in $n'\sub{COBE}$ 
can allow larger (smaller) 
values of $\omega_b$ ($\omega_{cdm}$).
The Hubble parameter $h$ is poorly constrained by CMB alone.

We have also considered the effect of variations in 
$n_{COBE}$ and $n'\sub{COBE}$ 
on the predicted level of rms
mass fluctuations in spheres of $8  h^{-1}\Mpc $.
At the moment, there is some sort of tension between the
possible values of this parameter obtained from
local cluster abundances ($\sigma_8 \sim 0.75$ for $\Omega_M=0.3$,
see e.g. \cite{seljakcluster})
and weak lensing ($\sigma_8 \sim 0.9$ for $\Omega_M=0.3$ see e.g.
\cite{refregier}).

\begin{figure}
\centering
\epsfysize=3.5cm 
\epsffile{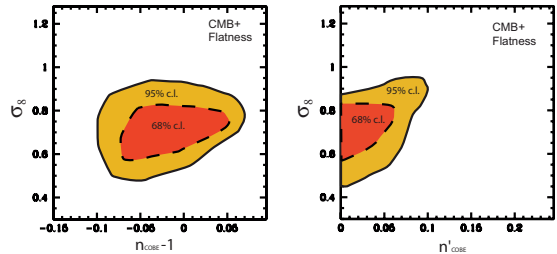} 
\caption[sc-fig2]{Likelihood contours in the $(n_{COBE}-1)-\sigma_8$ (Top) and
$n'\sub{COBE}-\sigma_8$
%$({\diff n\sub{COBE} }/{\diff \ln k})-\sigma_8$ 
planes.
}
\label{contours4}
\end{figure}

In Figure~\ref{contours4} we plot the likelihood contours 
from the CMB analysis for $\sigma_8$ against
$n_{COBE}-1$ (Top Panel) and $n'\sub{COBE}$
%${\diff n\sub{COBE} }/{\diff \ln k}$ 
(Bottom Panel) respectively.
As expected, increasing both $n_{COBE}-1$ 
and $n'\sub{COBE}$
%${\diff n\sub{COBE} }/{\diff \ln k}$ 
increases the
level of $\sigma_8$, but the correlation is not relevant enough
to actually detect a bending in the spectrum with 
present observations.

%[{\tt slightly expanded}]
We are in qualitative agreement with other recent constraints on 
a running spectral index
\cite{hannestad-run}. Precise comparison is difficult 
mainly because we did not consider a background
of gravity waves (that can weaken our final result) since
in the running-mass models this is %expected to be 
negligible.
Furthermore, we considered a more updated CMB dataset respect
to the one considered in \cite{hannestad-run} and
we considered different datasets other than CMB.
Finally,  \cite{hannestad-run} assumes that $n'$ 
can take any value, whereas the running-mass model
excludes significantly negative $n'$ (\eq{bound-dndlnk}).
The running-mass model also allows 
variation of $n'(k)$,  but this is not very
significant at the present level of observation.

%[{\tt bit about $V''''$}]
We end this discussion of our results by
 commenting on a recent analysis by Caprini, Hansen and Kunz
\cite{chk} which investigates the variation of $n'(k)$ that is allowed
by observation. First, they assume  that the flatness parameter
$\sigma^3$ is constant  to the end of slow-roll
inflation, and taking $N\sub{COBE}=50$, they
 find $\sigma^3 > -3.5\times 10^{-4}$.
The running-mass prediction \eq{sigma3} for $\sigma^3$
is roughly constant to the end of slow-roll inflation 
if $|c|\lsim 1/50$, and in this regime 
one needs $|s|\lsim 0.1$ to have $n$ sufficiently close to 1; combining
these results gives roughly $|\sigma^3|\lsim 4\times 10^{-4}$ in agreement
with \cite{chk}. However, \cite{chk} assume next that instead
$V''''$ is roughly constant, and derives a bound
$\mpl^4V''''/V>-0.02$. In contrast, the running-mass model predicts
$\mpl^4V''''/V=c(\mpl/\phi)^2$ which is exponentially large (and of
either sign) and strongly varying. Thus the bound on $V''''$ 
of \cite{chk}
does not apply to the running-mass model in any regime of its parameter
space. We remind the reader again that, at the moment,
 the running-mass model is the only well-motivated model giving 
significant running of the spectral index.

\medskip
\section{Conclusions}
\medskip

%[{\tt new para}]
The rather full analysis that we have described confirms the general
picture indicated by previous analyses \cite{covi}.
The allowed region in the $c$ vs $s$ plane depicted in 
Figure~\ref{contours}
%\ref{contours4}
should be compared with the region  shown in
Figure~\ref{laura101} which 
approximately delineates the
 theoretically disfavoured  region,
%because it requires some degree of fine-tuning, 
and also with the 
%bound
minimum value
$|c|\gsim 10^{-2}$ which is probably 
needed to generate enough running of the mass
even if we go from the Planck scale to $100\GeV$. Combining all of these,
%we see that only the version of the model with $c$ and $s$ both positive
%seems to be viable. In this regime, there is quite a long way to go before
%the data can exclude or confirm the model. 
we see that if $|c|$ is significantly above the minimum value,
only the version of the model with $c$ and $s$ both positive
is viable. In that case, the spectral index has significant running which
will be detectable in the forseeable future. On the other hand, 
if $|c|$ is really of order $10^{-2}$,
all choices of the signs of $c$ and $s$ are possible except maybe
negative $c$ with positive $s$. Furthermore, if that extreme case can be
realized in a viable running-mass model
the running of $n$  will be so small that it may never be detectable.

%In this paper we have presented new constraints on the 
%scale-dependence of the spectral index of the cosmological
%curvature perturbation.
Looking at the observational situation in more detail, our
 results show that the CMB data can put very strong
constraints on the value of the spectral index at large scales,
$n_{COBE} =1 + 2(s-c)$, but still allow a pretty large scale-dependence.
Other information on the power spectrum, like Lyman $\alpha$ data,
or strongest assumptions about the reionization epoch
are needed to reduce the parameter space in the $s+c$ direction.
Still values of $|c|$ of the order of 0.2, and therefore 
$n'\sub{COBE}$
%$\diff n\sub{COBE}/\diff \ln k$  
of the order $0.04$, 
are allowed.
Note also that our allowed region is more or less symmetric under 
reflection with respect to the $s+c=0$ line: this means that the present 
data are not sensitive enough to distinguish the variation of $n'$
that is predicted by the running-mass model.
%terms of higher
%order than the first in the expansion eq.~(\ref{taylorn}),
%and so cannot yet single out the shape of the spectral index 
%characteristic of the running mass models.
Finally, we show that the inclusion of the running in the
spectral index has a moderate effect on the present CMB
determination of the physical matter densities in
baryons $\omega_b$ and cold dark matter $\omega_{cdm}$.

%[{\tt the para about more general models removed as we are really
%focussing exclusively on the running mass model and this excludes
%all of the effects mentioned I believe}]
%As final remark, we should note that our CMB analysis was restricted 
%on a specific class of models with a limited numbers of parameters. 
%Including curvature, a gravity waves background, or removing our priors 
%on age would enlarge our C.L.. 
%Including drastical deviations from the standard
%model like topological defects or in the recombination scheme
% would drastically change the conclusions
%of our work. Each of these effect leaves a characteristic imprint on CMB,
%so hopefully with new data available in the near future it will be possible
%to severely constrain the scale-dependence of the spectral index of 
%the cosmological curvature perturbation.

\textit{Acknowledgements} 
It is a pleasure to thank Steen Hansen, 
Carolina Odman  and Joseph Silk for useful comments.
We acknowledge the use of CMBFAST~\cite{sz}.
AM is supported by PPARC.

\end{document}